\documentclass[runningheads]{llncs}

\usepackage{eccv}

\usepackage{eccvabbrv}

\usepackage{amsmath,amsfonts,amssymb}
\usepackage{multirow}
\usepackage{makecell}
\usepackage{float}
\usepackage{caption} % DO NOT CHANGE THIS AND DO NOT ADD ANY OPTIONS TO IT
\usepackage{enumitem}
\usepackage{comment}
\usepackage{graphicx}
\usepackage{booktabs}

\usepackage[accsupp]{axessibility}  % Improves PDF readability for those with disabilities.

\usepackage{hyperref}

\usepackage{orcidlink}

\begin{document}

\title{GET: Generative Embedding Translation for Medical Image Segmentation}
\titlerunning{Generative Embedding Translation}

\author{
\mbox{Md Maklachur Rahman}\inst{1} \and
\mbox{Md Hasan Al Banna}\inst{1} \and
\mbox{Saraf Anjum}\inst{2} \and
\mbox{Mahmudul Hasan}\inst{3} \and
\mbox{Tracy Hammond}\inst{1}
}

\authorrunning{M. M. Rahman et al.}

\institute{
Texas A\&M University, College Station, TX, USA
\and
Independent Researcher
\and
Alfa Laval\\
\email{\{maklachur, mdhasanalbanna, hammond\}@tamu.edu},\\
\email{sarafanjumeva@gmail.com},
\email{mahmudul.hasan@alfalaval.com}
}

\maketitle

\begin{abstract}
Generative segmentation provides an alternative to direct pixel-wise prediction by operating on learned latent representations, but effective image-to-mask translation must preserve target structure while remaining computationally efficient. We propose Generative Embedding Translation (GET), a structured embedding-translation framework that progressively transforms image embeddings into mask embeddings within the frozen latent space of a Stable Diffusion VAE. GET uses a U-Net-style Embedding Translation Network with 1.07M trainable parameters, combining Mobile Bottleneck Convolutions, Subsampled Self-Attention, and Multi-scale Feature Enrichment for local modeling, global context, and multi-scale refinement. Across five medical segmentation datasets, GET outperforms generative, CNN, and Transformer baselines. Compared with the strongest generative baseline, GMS, GET improves average Dice and IoU by 0.93\% and 1.26\%, reduces HD95 by 0.81 pixels, and uses 31.41\% fewer trainable parameters. Under bidirectional BUS--BUSI domain shift, GET further improves Dice and IoU by 3.51\% and 3.39\%, while reducing HD95 by 27.37 pixels. Our code is available at: \url{https://github.com/maklachur/GET}.

\end{abstract}

\section{Introduction}

Medical image segmentation is crucial for accurate diagnosis, treatment planning, and clinical decision-making. Traditional segmentation approaches predominantly use Convolutional Neural Networks (CNNs), such as U-Net \cite{unet}, UNet++ \cite{unet++}, AULUNet \cite{aulunet}, DeepLabV3+ \cite{deeplabv3}, and nnUNet \cite{nnunet}, which effectively capture hierarchical local features. However, CNN-based methods remain constrained in modeling long-range dependencies, which are important for segmenting complex anatomical structures and ambiguous boundaries \cite{oktay2018attention}.

Transformer-based models \cite{vit} address this limitation through global self-attention. Architectures such as TransUNet \cite{transunet}, SwinUNet \cite{swinunet}, and UNETR \cite{unetr} improve contextual modeling by capturing long-range interactions, but their attention mechanisms can introduce substantial computational cost. Their performance can also depend on large annotated datasets, which are often difficult to obtain in medical imaging \cite{trmmedicalsurvey,mambaliteunet}.

Generative segmentation provides a complementary direction by operating on learned latent representations rather than relying solely on direct pixel-wise prediction. Variational Autoencoders (VAEs) \cite{iso_kl} and Vector-Quantized VAEs (VQ-VAEs) \cite{taming} provide compact latent representations, while recent segmentation methods include diffusion-based approaches such as MedSegDiff-V2 \cite{medsegdiffv2} and SDSeg \cite{sdseg}, as well as direct latent-translation approaches such as GSS \cite{gss} and GMS \cite{gms}. Although prior work demonstrates the feasibility of translating image representations into mask representations, accurate segmentation requires the translation to preserve the spatial structure of the target. The transformation must recover the target's location, extent, shape, and boundaries from image appearance while remaining robust to variations in texture, contrast, and acquisition conditions. This motivates a structured translation process that can jointly model local detail, global context, and multi-scale spatial information while keeping the trainable component compact. A detailed discussion of related work is provided in Supplementary Material Section \ref{Supp-rel}.

Based on this perspective, we propose Generative Embedding Translation (GET), which formulates medical image segmentation as structured embedding translation within a frozen Stable Diffusion VAE (SD-VAE) space. The SD-VAE provides a fixed representation space for both images and masks, while a lightweight Embedding Translation Network (ETN) learns the task-specific transformation from image embeddings to mask embeddings. Rather than treating this transformation as a single mapping operation, ETN progressively translates the representation across multiple resolution stages, allowing spatial structure to be modeled and refined before the predicted mask embedding is decoded. Keeping the SD-VAE frozen further isolates the trainable segmentation component and limits the number of parameters that must be optimized.

The ETN is a five-stage U-Net-style translator that combines three complementary components. Mobile Bottleneck Convolution (MBC) provides efficient local feature extraction, Subsampled Self-Attention (SSA) captures global context at the bottleneck, and Multi-Scale Feature Enrichment (MFE) aggregates information across multiple receptive fields during decoding. Multi-stage latent reconstruction and additional L2 latent alignment constrain intermediate and final representations toward the target mask embedding, while Dice and Focal-Tversky losses directly supervise the decoded segmentation. Together, these components enable GET to preserve local structure, global organization, and multi-scale detail within a compact latent translation framework.

In summary, our main contributions are as follows:
\begin{itemize}

    \item We develop GET, a structured embedding-translation framework that progressively maps image embeddings to mask embeddings across multiple resolutions within a frozen SD-VAE space using a compact trainable translator.

    \item We design a multi-stage ETN with MBC, SSA, and MFE for efficient local modeling, global context aggregation, and multi-scale refinement, together with latent- and mask-space supervision.

    \item Experiments on BUS, BUSI, GlaS, HAM10000, and Kvasir--Instrument demonstrate consistent improvements over recent generative, CNN, and Transformer baselines. Compared with GMS, GET improves average Dice and IoU by 0.93\% and 1.26\%, reduces HD95 by 0.81 pixels, and uses 31.41\% fewer trainable parameters. Under bidirectional BUS--BUSI domain shift, GET further improves Dice and IoU by 3.51\% and 3.39\%, while reducing HD95 by 27.37 pixels.

\end{itemize}
\begin{figure*}[t]
  \begin{center}
  \includegraphics[width=0.99\textwidth]{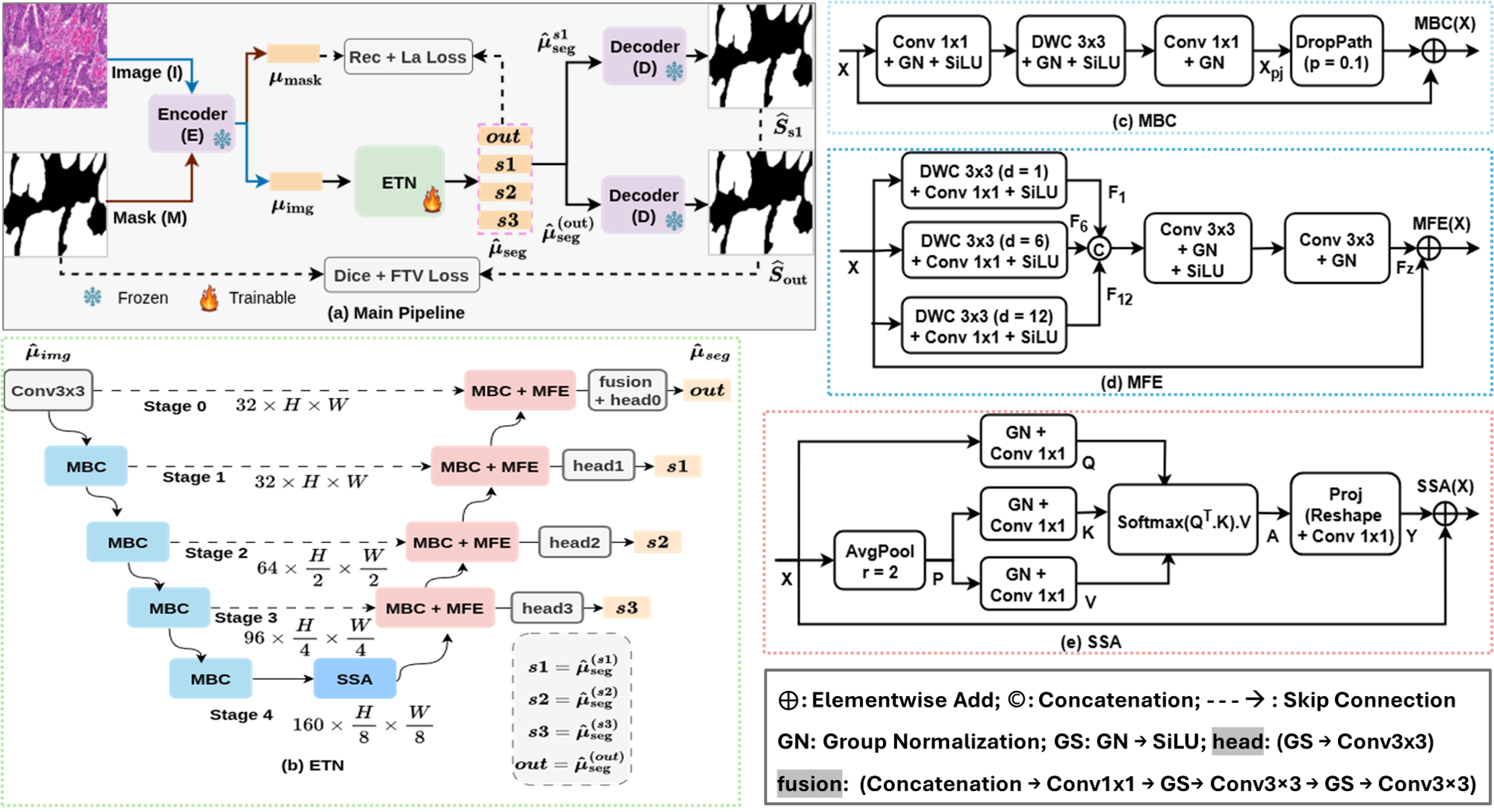}
  \end{center}
  \caption{Overview of the proposed framework.
  (a)~Main pipeline: a frozen SD-VAE encodes both image~($I$) and mask~($M$) into latent features~$\mu_{\text{img}}$ and~$\mu_{\text{mask}}$. 
  The lightweight ETN translates image latent $\mu_{\text{img}}$ to segmentation latent $\hat{\mu}_{\text{seg}}$. The \(s_1\), \(s_2\), \(s_3\), and \(out\) are the segmentation latents from different stages of ETN. The $\hat{\mu}_{\mathrm{seg}}^{(s_1)}$ and $\hat{\mu}_{\mathrm{seg}}^{(out)}$ denote the segmentation latent for \(s_1\) and \(out\), respectively, which are then decoded to generate the final masks.
  (b)~ETN integrates MBC, MFE, and SSA modules for latent translation.
  (c)~MBC acts as the encoder within ETN.
  (d)~MFE aggregates multi-scale features via parallel dilated branches in the decoder.
  (e)~SSA operates at the bottleneck, capturing long-range context through subsampled attention.}
  \label{fig:main_pipeline}
\end{figure*}

\section{Methodology}
\label{sec:methodology}
We formulate GET as structured embedding translation, where an image latent is progressively transformed into a segmentation latent within a frozen SD-VAE representation space. As shown in Figure~\ref{fig:main_pipeline}, our Generative Embedding Translation (GET) framework consists of two integrated components: (1) a frozen Stable Diffusion Variational Autoencoder (SD-VAE) encoder--decoder, where the encoder embeds images and masks into a shared semantic latent space and the decoder reconstructs masks from that space; and (2) a lightweight, trainable Embedding Translation Network (ETN) that maps image latents directly to segmentation latents. During inference, ETN predicts the segmentation latent from the input image, and the frozen SD-VAE decoder produces the final segmentation mask from this latent.

\subsection{Latent Embedding Formulation}
In this work, we employ the SD-VAE~\cite{stable_diffusion} as a frozen encoder-decoder pair \((E, D)\) to define a unified latent representation for both inputs and outputs.
Given an input image \(I\in\mathbb{R}^{3\times H'\times W'}\) and a ground-truth mask \(M^{\text{gt}}\in\mathbb{R}^{1\times H'\times W'}\), we first form a three-channel mask
\(M\in\mathbb{R}^{3\times H'\times W'}\) by replicating \(M^{\text{gt}}\) across channels.
Both \(I\) and \(M\) are then normalized to match the SD-VAE input domain, and passed through the same frozen encoder to obtain their latent embeddings:
\begin{equation}
    \mu_{\text{img}} = E(I), \qquad
    \mu_{\text{mask}} = E(M),
\end{equation}
where \(\mu_{\text{img}}, \mu_{\text{mask}} \in \mathbb{R}^{4\times H\times W}\) represent the image and mask latents, respectively, with \(H = H'/8\) and \(W = W'/8\).

As SD-VAE is pre-trained on large-scale natural images, it provides a high-fidelity, semantically structured latent manifold that transfers well to medical imagery, enabling our generative segmentation model to train without additional backbone adaptation~\cite{gms,gss}.

Our ETN, denoted by \(\mathcal{T}\), learns a direct mapping from the image latent \(\mu_{\text{img}}\) to a predicted segmentation latent \(\hat{\mu}_{\text{seg}}\), which is decoded by the frozen SD-VAE decoder \(D\):
\begin{equation}
    \hat{\mu}_{\text{seg}} = \mathcal{T}(\mu_{\text{img}}), \qquad
    \widetilde{S} = D(\hat{\mu}_{\text{seg}}), \qquad
    \widehat{S} = \Pi(\widetilde{S}).
\end{equation}
Here, \(\hat{\mu}_{\text{seg}} \in \mathbb{R}^{4\times H\times W}\) denotes the predicted latent representation, \(\widetilde{S} \in \mathbb{R}^{3\times H'\times W'}\) is the three-channel mask reconstruction decoded by the frozen SD-VAE, and \(\Pi(\cdot)\) is a fixed projection that converts \(\widetilde{S}\) into a single-channel segmentation map. Specifically, we average the decoded channels after mask-domain normalization to obtain \(\widehat{S} \in \mathbb{R}^{1\times H'\times W'}\), which is thresholded at \(0.5\) during evaluation to produce the final binary mask. 
By operating entirely in the latent space and keeping \((E,D)\) frozen, GET substantially reduces trainable parameters and computation while leveraging the generative prior of SD-VAE for stable and structurally coherent segmentation.

\subsection{Embedding Translation Network (ETN)}
The ETN is a five-stage U-Net–style latent translator with four encoder–decoder stages and a central bottleneck, operating entirely in the SD-VAE latent space. 
Starting from the image latent \(\mu_{\mathrm{img}}\), the encoder gradually expands feature channels to \(\{32, 64, 96, 160\}\) using MBC blocks, while the bottleneck applies SSA over the coarsest latent maps to capture global context. 
The decoder then symmetrically upsamples and fuses encoder features using MBC and MFE, and a final fusion layer aggregates encoder and decoder output to the predicted segmentation latent \(\hat{\mu}_{\mathrm{seg}}\). Between ETN stages, we use $2\times$ average pooling for downsampling and nearest-neighbor interpolation for upsampling, while auxiliary heads use GN--SiLU--$3\times3$ convolution to predict four-channel latents.

\subsubsection{Mobile Bottleneck Convolution (MBC)}

The MBC module serves as a building block in both the encoder and decoder of ETN. It enhances feature representations with minimal computational overhead by first projecting the input tensor \( X \in \mathbb{R}^{B\times C\times H\times W} \) into a higher-dimensional space via a \(1\times1\) convolution, Group Normalization (GN), and SiLU activation. This is followed by a \(3\times3\) depthwise convolution ($\mathrm{DWC}$) to capture local spatial patterns, again followed by GN and SiLU. These operations can be expressed as:
\begin{equation}
    X_{\mathrm{pj}} = \mathcal{G}\mathcal{S}\big(\mathrm{DWC}_{3\times3}(\mathcal{G}\mathcal{S}(\mathrm{Conv}_{1\times1}(X)))\big).
\end{equation}

Here, each \(\mathrm{GN\rightarrow SiLU}\) pair ($\mathcal{G}\mathcal{S}$) is applied sequentially after the corresponding convolution layer. Finally, we project the output back to the original channel dimension using another \(1\times1\) convolution followed by GN, and add a residual connection with drop-path (DP) $(p=0.1)$ regularization to stabilize training and encourage feature reuse:
\begin{equation}
\mathrm{MBC}(X) = X + \mathrm{DP}\big(\mathrm{GN}(\mathrm{Conv}_{1\times1}(X_{\mathrm{pj}}))\big).
\end{equation}

\subsubsection{Multi-Scale Feature Enrichment (MFE)}
Our MFE module enhances decoder features by aggregating multi‑scale context in a single, lightweight block. Given an input \( X \in \mathbb{R}^{B\times C\times H\times W} \), it branches into three parallel \(3\times3\) depthwise convolutions with dilation rates \( d \in \{1,6,12\} \), each followed by a \(1\times1\) pointwise convolution and SiLU:
\begin{equation}
F_r = \mathrm{SiLU}\bigl(\mathrm{Conv}_{1\times1}(\mathrm{DWC}_{3\times3}^{(d)}(X))\bigr),
\end{equation}
where  \( d \in \{1,6,12\} \). Inspired by DeepLabv3+ \cite{deeplabv3}, varying \( d \) allows each branch to capture fine details (\(d{=}1\)) or broader semantic information (\(d{=}6,12\)). 
The feature maps are concatenated and processed through two $3\times3$ convolutions with GroupNorm and SiLU activation, followed by a skip connection that adds back the input:
\begin{equation}
\begin{gathered}
F_{\mathrm{z}} =
\mathrm{GN}\!\left(
    \mathrm{Conv}_{3\times3}\!\left(
        \mathcal{G}\mathcal{S}\!\left(
            \mathrm{Conv}_{3\times3}([F_1, F_6, F_{12}])
        \right)
    \right)
\right),\\[3pt]
\mathrm{MFE}(X) = X + F_{\mathrm{z}},
\end{gathered}
\end{equation}
which ensures the combined representation of local details and long-range context before the next upsampling step.

\subsubsection{Subsampled Self-Attention (SSA)}
Our SSA module integrates global context at the bottleneck by computing attention between full-resolution queries and subsampled keys/values, reducing complexity by about \(r^2\).
Given an input
$X \in \mathbb{R}^{B\times C\times H\times W}$,
we first normalize and project it to obtain full‑resolution queries:
\begin{equation}
Q = \mathrm{Conv}_{1\times1}\bigl(\mathrm{GN}(X)\bigr)\;\in\;\mathbb{R}^{B\times C\times H\times W}.
\end{equation}
% where $\mathrm{GN}$ is GroupNorm and $W_q$ is a $1\times1$ convolution. 
We select the number of heads $h$ as the largest divisor of $C$ from $\{8,6,4,3,2,1\}$, giving a per‑head dimension $d = C/h$.
Next, we downsample spatially by a factor \( r=2 \) (see Supplementary Material Section \ref{supp:r_effect}) via average pooling to form
$P = \mathrm{AvgPool}_r(X)\;\in\;\mathbb{R}^{B\times C\times H/r\times W/r}$,
and normalize, then project to produce keys and values:
\begin{equation}
K = \mathrm{Conv}_{1\times1}\bigl(\mathrm{GN}(P)\bigr),\quad
V = \mathrm{Conv}_{1\times1}\bigl(\mathrm{GN}(P)\bigr),
\end{equation}
each in $\mathbb{R}^{B\times C\times H/r\times W/r}$. We then reshape into multi‑head tensors of shape $(B,h,d,N)$ for $Q$ with $N=H\,W$, and $(B,h,d,M)$ for $K,V$ with $M=(H\,W)/r^2$, and compute scaled dot‑product attention:
\begin{equation}
A = \mathrm{softmax}_{\text{col}}\!\bigl((Q^\top K)/\sqrt{d}\bigr), \qquad
Z = A\,V^\top,
\end{equation}
where $Q^\top\!:\!(B,h,d,N)\!\to\!(B,h,N,d)$,  
$A\in\mathbb{R}^{B\times h\times N\times M}$, \(V^{\top}\!:\!(B,h,d,M)\!\to\!(B,h,M,d)\),
and $Z\in\mathbb{R}^{B\times h\times N\times d}$. We reassemble the heads to produce $Y\in\mathbb{R}^{B\times C\times H\times W}$  
from $Z$ and finally project and add a residual to obtain $\mathrm{SSA}(X)=X+\mathrm{Conv}_{1\times1}(Y)$.
This reduces complexity from $O\bigl((HW)^2\bigr)$ to 
$O\bigl(HW \times (HW/r^2)\bigr)$, enabling efficient global‐context integration.

\subsection{Training Objective and Loss Function}
\label{sec:loss}

We train the ETN to translate the image latent ($\mu_{\mathrm{img}}$) into the segmentation latent ($\hat{\mu}_{\mathrm{seg}}$) such that, when decoded by the frozen SD-VAE, it produces a high-quality segmentation mask. The overall training objective integrates multiple losses: $\mathcal{L}_{\mathrm{rec}}$ for latent reconstruction, $\mathcal{L}_{\mathrm{la}}$ for L2 latent alignment, and $\mathcal{L}_{\mathrm{dice}}+\mathcal{L}_{\mathrm{ftv}}$ for supervision in the mask space.

\subsubsection{Latent Reconstruction Loss}
To supervise latent reconstruction, we apply an $L_1$ loss~\cite{l1loss} between the predicted latent $\hat{\mu}_{\mathrm{seg}}^{(k)}$ and the mask latent $\mu_{\mathrm{mask}}$ at each deep supervision (DS) head $k$:
\begin{equation}
\mathcal{L}_{\mathrm{rec}}
=
\sum_{k \in \{\mathrm{out},s1,s2,s3\}}
w_k
\left\|
\hat{\mu}_{\mathrm{seg}}^{(k)}
-
\mu_{\mathrm{mask}}
\right\|_1.
\label{eq:main_lrec}
\end{equation}

For auxiliary predictions at different resolutions, we resize $\mu_{\mathrm{mask}}$ to the prediction resolution before computing the $L_1$ loss. We use $(w_{\mathrm{out}},w_{s1},w_{s2},w_{s3})=(1.0,0.4,0.3,0.2)$ based on experiments (see Supplementary Material Section~\ref{supp:ds_effect}).

\subsubsection{L2 Latent Alignment}
In addition to the multi-stage L1 reconstruction objective, we apply an L2 loss to the final predicted latent to further align it with the ground-truth mask latent:
\begin{equation}
\mathcal{L}_{\mathrm{la}}
=
\frac{\lambda_{\mathrm{la}}}{2}
\left\|
\hat{\mu}_{\mathrm{seg}}^{(\mathrm{out})}
-
\mu_{\mathrm{mask}}
\right\|_2^2,
\label{eq:latent_alignment}
\end{equation}
where $\lambda_{\mathrm{la}}=0.01$ is selected experimentally (see Supplementary Material Section~7.4). This term complements the multi-stage L1 reconstruction objective by providing an additional distance constraint between the final predicted latent and the target mask latent.

\subsubsection{Segmentation Loss in Mask Space}
We use a combination of Dice \cite{vnet} and Focal-Tversky (FTV) \cite{ftvloss} losses for mask-space supervision $\mathcal{L}_{\mathrm{seg}}$, where Dice loss helps to maximize overall region overlap, while standard Focal-Tversky $(\alpha, \beta, \gamma) = (0.7, 0.3, 0.75)$ focuses learning on hard, imbalanced pixels (see Supplementary Material Section \ref{supp:train_loss_formulas}). We define it as:

\begin{equation}
\mathcal{L}_{\mathrm{seg}}
= \big(\mathcal{L}_{\mathrm{dice}} + \mathcal{L}_{\mathrm{ftv}}\big)(\widehat{S}_{\mathrm{out}}, M)
+ \lambda \big(\mathcal{L}_{\mathrm{dice}} + \mathcal{L}_{\mathrm{ftv}}\big)(\widehat{S}_{\mathrm{s1}}, M).
\label{eq:main_lseg}
\end{equation}
where $\widehat{S}_{\mathrm{out}}$, $\widehat{S}_{\mathrm{s1}}$ are the decoder generated segmentation for final output latent ($\hat{\mu}_{\mathrm{seg}}^{(out)}$) and auxiliary latent $\hat{\mu}_{\mathrm{seg}}^{(s1)}$, respectively. $M$ is the ground-truth mask. We experimentally set $\lambda = 0.5$ (see Supplementary Material Section \ref{supp:aux_head_effect}) for the $\mathrm{s1}$ head to balance fine-structure guidance and mask consistency, enhancing boundary sharpness while minimizing noise.
This combined loss enables the model to generate accurate masks even under severe class imbalance~\cite{losssurvey}.
% This combined loss helps our model produce an accurate mask even under severe class imbalance \cite{losssurvey}.
Therefore, we can summarize our total loss as:

\begin{equation}
    \mathcal{L}_{\mathrm{total}} = 
    \mathcal{L}_{\mathrm{rec}}
    + \mathcal{L}_{\mathrm{la}} 
    + \mathcal{L}_{\mathrm{seg}}.
\end{equation}

\newcommand{\meansd}[2]{#1\textsuperscript{#2}}
\setlength{\tabcolsep}{1mm}
\renewcommand{\arraystretch}{1.02}

\begin{table*}[t]
\footnotesize
\begin{center}
\resizebox{\linewidth}{!}{
\begin{tabular}{c|l|c||c|c|c||c|c|c}
\hline
\multirow{2}{*}{AT} & \multirow{2}{*}{Model} & \multirow{2}{*}{\begin{tabular}[c]{@{}c@{}}Trainable\\Params (M)$\downarrow$\end{tabular}} & \multicolumn{3}{c|}{BUS} & \multicolumn{3}{c}{BUSI} \\
\cline{4-9}
 &  &  & DSC$\uparrow$ & IoU$\uparrow$ & HD95$\downarrow$ & DSC$\uparrow$ & IoU$\uparrow$ & HD95$\downarrow$ \\
\hline
 C & UNet~\cite{unet} & 14.03   & \meansd{81.65}{.50} & \meansd{70.98}{.73} & \meansd{17.50}{1.90} & \meansd{72.42}{.79} & \meansd{63.18}{1.05} & \meansd{35.22}{4.10} \\
 C & MultiResUNet \cite{ibtehaz2020multiresunet} & 7.26   & \meansd{80.56}{.47} & \meansd{70.54}{.69} & \meansd{19.04}{2.02} & \meansd{72.58}{.82} & \meansd{62.77}{1.08} & \meansd{33.99}{3.85} \\
 C & ACC-UNet \cite{accunet} & 16.77   & \meansd{83.30}{.51} & \meansd{73.36}{.77} & \meansd{16.62}{1.73} & \meansd{77.34}{.65} & \meansd{68.71}{.92} & \meansd{25.26}{2.85} \\
 C & nnUNet \cite{nnunet} & 20.6   & \meansd{85.61}{.36} & \meansd{78.53}{.58} & \meansd{11.56}{1.15} & \meansd{79.60}{.59} & \meansd{71.20}{.86} & \meansd{21.89}{2.41} \\
 C & EGE-UNet$\star$ \cite{egeunet} & 0.05   & \meansd{72.94}{.75} & \meansd{62.15}{.98} & \meansd{27.57}{3.05} & \meansd{75.32}{.89} & \meansd{60.42}{1.14} & \meansd{29.30}{3.42} \\
 C & LB-UNet$\star$ \cite{lbunet} & 0.04   & \meansd{72.42}{.78} & \meansd{61.50}{1.02} & \meansd{28.13}{3.18} & \meansd{74.76}{.88} & \meansd{59.70}{1.12} & \meansd{30.11}{3.55} \\
 T & SwinUNet \cite{swinunet} & 31.32   & \meansd{80.27}{.53} & \meansd{69.61}{.76} & \meansd{20.61}{2.14} & \meansd{76.21}{.75} & \meansd{66.30}{1.03} & \meansd{28.46}{3.14} \\
 T & SME-SwinUNet \cite{wang2022smeswin} & 169.83   & \meansd{78.87}{.64} & \meansd{67.28}{.88} & \meansd{22.16}{2.36} & \meansd{73.98}{.89} & \meansd{62.82}{1.18} & \meansd{30.33}{3.35} \\
 T & UCTransNet \cite{uctransnet} & 65.64   & \meansd{83.59}{.49} & \meansd{73.96}{.74} & \meansd{16.14}{1.66} & \meansd{76.70}{.75} & \meansd{67.70}{1.06} & \meansd{25.23}{2.86} \\
 G & MedSegDiff-V2 \cite{medsegdiffv2} & 129.43   & \meansd{83.38}{.67} & \meansd{74.58}{1.02} & \meansd{16.83}{1.79} & \meansd{71.22}{.93} & \meansd{62.61}{1.23} & \meansd{38.61}{4.63} \\
 G & SDSeg \cite{sdseg} & 329.17   & \meansd{82.62}{.60} & \meansd{73.67}{.90} & \meansd{20.34}{2.08} & \meansd{72.66}{.91} & \meansd{63.40}{1.21} & \meansd{36.93}{4.34} \\
 G & GSS \cite{gss} & 49.84   & \meansd{84.76}{.53} & \meansd{77.43}{.83} & \meansd{22.55}{2.30} & \meansd{79.71}{.66} & \meansd{71.43}{.97} & \meansd{27.96}{3.19} \\
 G & GMS \cite{gms} & 1.56   & \meansd{88.57}{.48} & \meansd{80.80}{.78} & \meansd{6.59}{.66} & \meansd{81.33}{.77} & \meansd{72.44}{1.15} & \meansd{19.66}{2.16} \\ 

G & \textbf{GET (Ours)} & \textbf{1.07}   
& \meansd{\textbf{90.76}}{.42} & \meansd{\textbf{83.48}}{.70} & \meansd{\textbf{4.84}}{.46} 
& \meansd{\textbf{83.03}}{.71} & \meansd{\textbf{74.56}}{1.08} & \meansd{\textbf{16.99}}{1.90} \\
\hline
\end{tabular}
 }
\end{center}
\caption{Quantitative results on BUS and BUSI datasets. AT denotes the primary architecture type: C = CNN, T = Transformer, G = Generative. $\uparrow$ indicates higher is better, $\downarrow$ lower is better, and $\star$ marks models with fewer parameters than ours. All SOTA baselines are reproduced using their public code and papers, and trained and tested on the same train--test splits as our model. Results are averaged over five independent runs and reported as mean$\pm$SD. Best scores are shown in bold.}
\label{tab:quantitative_bus_busi}
\end{table*}

\section{Experiments and Results}
\subsection{Datasets, Implementation, and Evaluation}

We evaluate GET on five medical segmentation datasets across different imaging modalities, following the preparation protocols of prior works~\cite{gms, matchseg}. BUS~\cite{bus} contains 163 breast ultrasound images (132 train / 31 test), while BUSI~\cite{busi} includes 647 ultrasound scans (517 train / 130 test) with lesion masks under low contrast and speckle noise. GlaS~\cite{glas} comprises 165 histology images (85 train / 80 test) with gland annotations. HAM10000~\cite{ham10000} contains 10015 dermoscopic RGB images with binary lesion masks~\cite{ham10000_masks} (8015 train / 2000 test). Kvasir-Instrument~\cite{kvasir_instrument} contains 590 endoscopic frames (472 train / 118 test) annotated for surgical tools under varying illumination and occlusion. All images and masks are resized to $224\times224$, normalized, and augmented with flips, rotations, and jitter during training.

We implement GET in PyTorch and train the ETN for 800 epochs on an NVIDIA RTX 3090 Ti GPU with 24 GB VRAM and batch size 8. The SD-VAE encoder--decoder~\cite{stable_diffusion} remains frozen, while only the ETN is optimized using AdamW~\cite{adamw} with learning rate 0.002, weight decay $10^{-4}$, and cosine annealing with warm restarts~\cite{cosineannealing}. Training uses mixed precision, TF32 acceleration, and an exponential moving average with $\tau=0.999$. Following the evaluation protocol of prior works~\cite{gms, matchseg}, we use the predefined train--test splits without a separate validation set and select the EMA checkpoint with the highest Dice on the designated test split. Thus, the designated test split also serves as a development split for checkpoint selection, while all compared baselines are evaluated under the same protocol.

We report Dice Similarity Coefficient (DSC) and Intersection over Union (IoU) in percent for overlap accuracy, and 95th percentile Hausdorff Distance (HD95) in pixels for boundary accuracy. Performance is evaluated on the designated test partitions using the same data splits across methods. Metric definitions are provided in Supplementary Material Section~\ref{supp:eval_metrics}.

\subsection{Comparison with SOTA Methods} 
We compare GET with CNN-based (UNet, MultiResUNet, ACC-UNet, nnUNet, EGE-UNet, and LB-UNet), Transformer-based (UCTransNet, SME-SwinUNet, and SwinUNet), and recent generative methods (MedSegDiff-V2, SDSeg, GSS, and GMS) across five benchmarks (Tables~\ref{tab:quantitative_bus_busi},~\ref{tab:quantitative_glas_kvasir_ham}). 

On the ultrasound datasets (BUS, BUSI), GET achieves DSC scores of 90.76\% and 83.03\%, outperforming GMS by 2.19 and 1.70 percentage points, respectively, while reducing HD95 from 6.59 to 4.84 pixels on BUS and from 19.66 to 16.99 pixels on BUSI. On HAM10000, GET achieves 94.23\% DSC, 89.95\% IoU, and 9.06 pixels HD95, improving over GMS in all three metrics. On the histology and endoscopy datasets (GlaS, Kvasir--Instrument), GET achieves the best DSC and IoU, with a modest HD95 improvement on GlaS and a small trade-off on Kvasir--Instrument. Overall, GET achieves these gains with only 1.07M trainable parameters, substantially fewer than most competing methods. Additionally, we present the model's Efficiency-Accuracy Trade-off comparison in the Supplementary Material, Section \ref{supp:model_eff_trade-off}.

\begin{figure*}[t]
  \begin{center}
  \includegraphics[width=\textwidth]{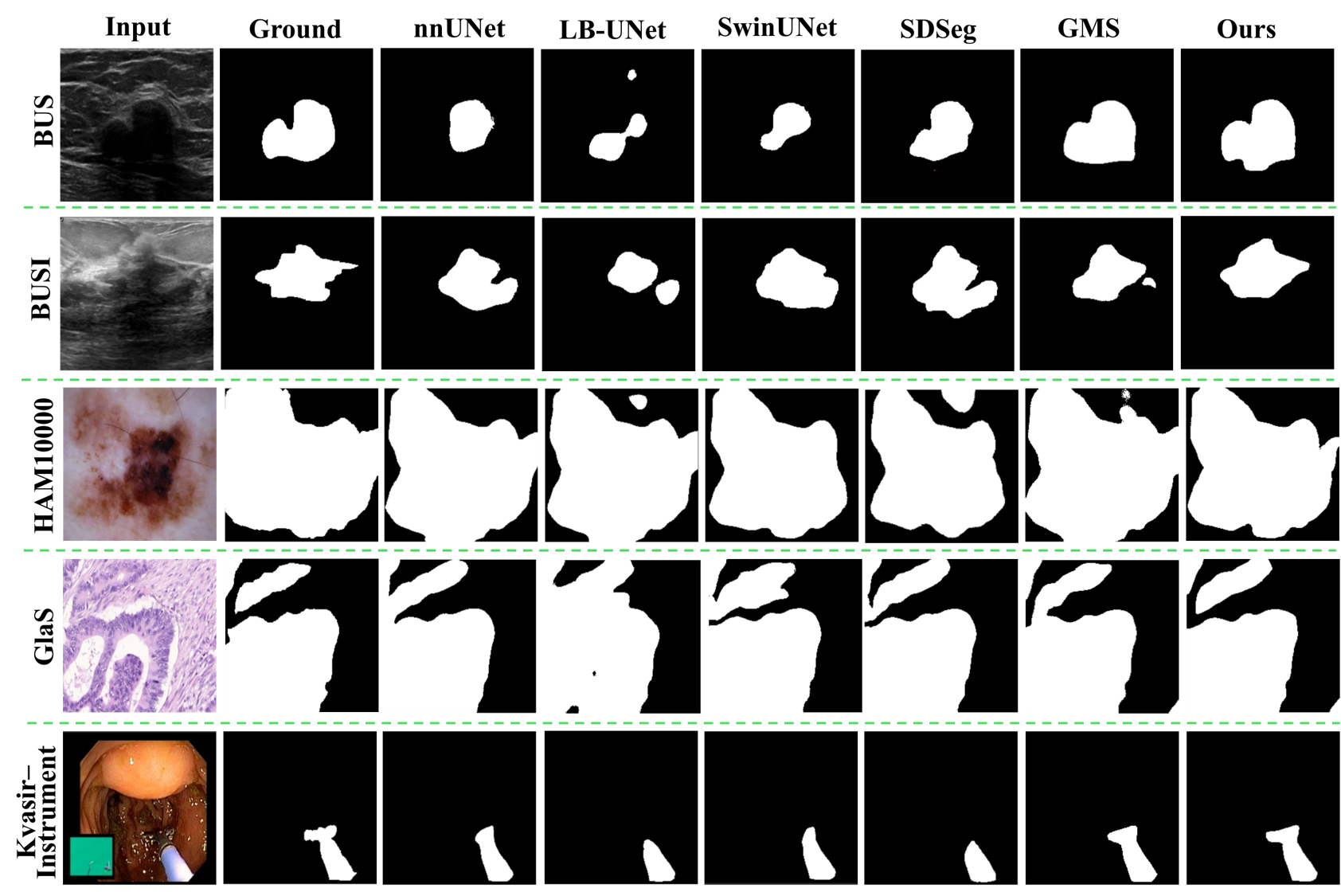}
  \end{center}
  \vspace{10pt}
\caption{Qualitative comparison on BUS, BUSI, HAM10000, GlaS, and Kvasir--Instrument. We show the input image, ground-truth mask, and predictions from representative segmentation methods across the five datasets.}
  \label{fig:qc}
\end{figure*}

Figure~\ref{fig:qc} presents qualitative comparisons across all five datasets, showing the ground-truth masks and predictions from representative segmentation methods.

\begin{table*}[h]
\footnotesize
\setlength{\tabcolsep}{.5mm}
\renewcommand{\arraystretch}{1.05}
\begin{center}
\resizebox{\linewidth}{!}{
\begin{tabular}{l|ccc|ccc|ccc}
\hline
\multirow{2}{*}{Model}  
& \multicolumn{3}{c|}{GlaS} 
& \multicolumn{3}{c|}{Kvasir--Instrument} 
& \multicolumn{3}{c}{HAM10000} \\ 
\cline{2-10}
 & DSC$\uparrow$ & IoU$\uparrow$ & HD95$\downarrow$ 
 & DSC$\uparrow$ & IoU$\uparrow$ & HD95$\downarrow$
 & DSC$\uparrow$ & IoU$\uparrow$ & HD95$\downarrow$ \\
\hline
UNet              
& \meansd{88.08}{.47} & \meansd{80.17}{.72} & \meansd{18.27}{1.96} 
& \meansd{93.86}{.29} & \meansd{89.30}{.49} & \meansd{8.69}{.79} 
& \meansd{92.36}{.29} & \meansd{87.14}{.50} & \meansd{13.51}{1.22} \\

MultiResUNet      
& \meansd{88.41}{.50} & \meansd{80.45}{.77} & \meansd{17.29}{1.83} 
& \meansd{92.33}{.27} & \meansd{87.06}{.46} & \meansd{9.48}{.86} 
& \meansd{92.64}{.26} & \meansd{87.43}{.46} & \meansd{13.29}{1.17} \\

ACC-UNet          
& \meansd{88.72}{.46} & \meansd{81.03}{.70} & \meansd{16.92}{1.69} 
& \meansd{93.96}{.31} & \meansd{89.82}{.57} & \meansd{8.71}{.76} 
& \meansd{93.33}{.25} & \meansd{88.67}{.45} & \meansd{10.58}{.92} \\

nnUNet            
& \meansd{87.30}{.43} & \meansd{78.32}{.68} & \meansd{19.98}{2.12} 
& \meansd{93.97}{.26} & \meansd{90.24}{.45} & \meansd{8.50}{.72} 
& \meansd{93.89}{.28} & \meansd{89.43}{.51} & \meansd{9.31}{.79} \\

EGE-UNet$\star$   
& \meansd{83.20}{.67} & \meansd{71.24}{.99} & \meansd{28.87}{2.98} 
& \meansd{92.62}{.38} & \meansd{86.25}{.68} & \meansd{9.06}{.87} 
& \meansd{93.03}{.31} & \meansd{87.97}{.54} & \meansd{11.14}{.98} \\

LB-UNet$\star$    
& \meansd{82.33}{.62} & \meansd{71.28}{.94} & \meansd{28.82}{3.11} 
& \meansd{92.60}{.34} & \meansd{86.21}{.62} & \meansd{9.07}{.93} 
& \meansd{93.80}{.37} & \meansd{88.32}{.66} & \meansd{10.30}{.88} \\

SwinUNet          
& \meansd{86.52}{.57} & \meansd{77.01}{.91} & \meansd{19.49}{1.99} 
& \meansd{92.00}{.40} & \meansd{85.80}{.70} & \meansd{9.16}{.88} 
& \meansd{93.58}{.35} & \meansd{88.80}{.63} & \meansd{10.32}{.91} \\

SME-SwinUNet      
& \meansd{83.81}{.66} & \meansd{72.90}{.95} & \meansd{26.08}{2.74} 
& \meansd{93.35}{.33} & \meansd{88.32}{.60} & \meansd{8.89}{.80} 
& \meansd{92.81}{.34} & \meansd{87.28}{.60} & \meansd{12.34}{1.07} \\

UCTransNet        
& \meansd{87.23}{.52} & \meansd{78.89}{.80} & \meansd{20.68}{2.06} 
& \meansd{93.31}{.35} & \meansd{88.55}{.63} & \meansd{8.82}{.82} 
& \meansd{93.50}{.37} & \meansd{88.82}{.66} & \meansd{10.92}{.95} \\

MedSegDiff-V2     
& \meansd{86.72}{.55} & \meansd{76.89}{.82} & \meansd{20.14}{2.09} 
& \meansd{92.31}{.36} & \meansd{87.24}{.66} & \meansd{9.05}{.83} 
& \meansd{91.93}{.41} & \meansd{87.12}{.72} & \meansd{13.19}{1.16} \\

SDSeg             
& \meansd{86.68}{.60} & \meansd{76.11}{.97} & \meansd{21.55}{2.25} 
& \meansd{91.24}{.42} & \meansd{86.56}{.73} & \meansd{9.37}{.90} 
& \meansd{92.47}{.43} & \meansd{87.41}{.76} & \meansd{12.42}{1.06} \\

GSS               
& \meansd{87.52}{.53} & \meansd{79.34}{.85} & \meansd{18.61}{1.88} 
& \meansd{93.71}{.37} & \meansd{89.24}{.65} & \meansd{7.22}{.64} 
& \meansd{93.07}{.46} & \meansd{88.24}{.82} & \meansd{10.88}{.98} \\

GMS               
& \meansd{88.93}{.48} & \meansd{81.08}{.83} & \meansd{16.41}{1.57} 
& \meansd{94.28}{.32} & \meansd{90.09}{.58} & \textbf{\meansd{7.00}{.56}} 
& \meansd{94.06}{.40} & \meansd{89.58}{.71} & \meansd{9.34}{.83} \\ 

\textbf{GET (Ours)} 
& \meansd{\textbf{89.24}}{.44} & \meansd{\textbf{81.67}}{.76} & \meansd{\textbf{15.62}}{1.45}
& \meansd{\textbf{94.55}}{.35} & \meansd{\textbf{90.63}}{.62} & \meansd{8.45}{.71}
& \meansd{\textbf{94.23}}{.30} & \meansd{\textbf{89.95}}{.55} & \meansd{\textbf{9.06}}{.82} \\
\hline
\end{tabular}
}
\end{center}

\caption{Results on GlaS, Kvasir--Instrument, and HAM10000 datasets. All SOTA baselines are reproduced using their public code and trained/tested on the same splits as ours. Results are averaged over five runs and reported as mean$\pm$SD.}
\label{tab:quantitative_glas_kvasir_ham}
\end{table*}

\subsection{Cross-Domain Generalization}
To evaluate robustness across clinical settings, we perform bidirectional cross-domain experiments between BUS and BUSI. Although both are breast ultrasound datasets, they originate from different acquisition settings, creating shifts in texture, contrast, and noise. We train on the BUS training set and test on BUSI, and vice versa (Table~\ref{tab:cross_domain_new}). For BUS$\rightarrow$BUSI, GET achieves 64.63\% DSC and 54.69\% IoU, outperforming GMS by 2.91 percentage points in DSC. For BUSI$\rightarrow$BUS, GET reaches 84.39\% DSC and 76.38\% IoU, improving over GMS by 4.11 and 6.34 percentage points, respectively, while reducing HD95 by 6.75 pixels. The consistent gains in both directions suggest that the learned embedding translation remains effective despite changes in image appearance and acquisition characteristics. These results demonstrate stronger robustness under cross-center domain shift.

\begin{table}[t]
\footnotesize
\begin{center}
\begin{tabular}{l|ccc|ccc}
\hline
\multirow{2}{*}{Model}
  & \multicolumn{3}{c|}{BUSI $\to$ BUS}
  & \multicolumn{3}{c}{BUS $\to$ BUSI} \\ 
\cline{2-4}\cline{5-7}
 & DSC$\uparrow$ & IoU$\uparrow$ & HD95$\downarrow$
 & DSC$\uparrow$ & IoU$\uparrow$ & HD95$\downarrow$\\

\hline
UNet            & 63.20 & 51.11 & 46.50 & 53.71 & 47.81 & 98.07 \\
MultiResUNet    & 61.61 & 50.25 & 53.71 & 56.24 & 50.32 & 92.40 \\
ACC‑UNet        & 64.71 & 52.39 & 42.08 & 47.61 & 46.68 & 136.11 \\
nnUNet          & 78.33 & 67.84 & 20.90 & 59.18 & 53.75 & 88.27 \\
EGE‑UNet$\star$ & 68.90 & 56.81 & 33.94 & 54.37 & 50.39 & 106.08 \\
LB-UNet$\star$  & 73.08 & 61.54 & 27.46 & 57.98 & 53.19 & 94.06 \\
SwinUNet        & 78.38 & 67.99 & 21.29 & 57.49 & 51.76 & 90.56 \\
SME‑SwinUNet    & 74.68 & 63.18 & 24.01 & 58.39 & 52.90 & 89.71 \\
UCTransNet      & 72.82 & 61.58 & 29.40 & 56.73 & 51.22 & 93.29 \\
MedSegDiff‑V2   & 69.74 & 57.51 & 31.32 & 55.03 & 49.88 & 98.83 \\
SDSeg           & 73.90 & 62.55 & 26.49 & 56.89 & 51.98 & 95.92 \\
GSS             & 68.85 & 57.39 & 37.11 & 58.58 & 53.63 & 91.99 \\
GMS             & 80.28 & 70.04 & 17.82 & 61.72 & 54.25 & 84.28 \\ 
\textbf{GET (Ours)}
                 & \textbf{84.39} & \textbf{76.38} & \textbf{11.07}
                 & \textbf{64.63} & \textbf{54.69} & \textbf{36.30} \\
\hline
\end{tabular}
\end{center}

\caption{Cross-domain generalization results between BUS and BUSI. BUSI$\rightarrow$BUS denotes training on BUSI and testing on BUS, while BUS$\rightarrow$BUSI denotes training on BUS and testing on BUSI. Best results are highlighted in bold.}
\label{tab:cross_domain_new}
\end{table}

\begin{table*}[!t]
\footnotesize
\begin{center}
\begin{tabular}{ccc|ccc|ccc|ccc}
\hline
\multicolumn{3}{c|}{Modules} 
& \multicolumn{3}{c|}{BUS} 
& \multicolumn{3}{c|}{BUSI} 
& \multicolumn{3}{c}{GlaS} \\
\cline{1-3} \cline{4-12}
MBC & MFE & SSA &
DSC$\uparrow$ & IoU$\uparrow$ & HD95$\downarrow$
& DSC$\uparrow$ & IoU$\uparrow$ & HD95$\downarrow$
& DSC$\uparrow$ & IoU$\uparrow$ & HD95$\downarrow$ \\
\hline \hline
\multicolumn{3}{c|}{Baseline}            & 88.48 & 80.05 & 6.72 & 79.23 & 70.73 & 21.06 & 87.38 & 78.56 & 18.82 \\ \hline
\checkmark   &              &              & 89.34 & 81.26 & 6.53 & 80.27 & 71.33 & 19.43 & 87.68 & 79.24 & 17.84 \\
             & \checkmark   &              & 89.15 & 80.95 & 6.76 & 81.01 & 72.07 & 21.15 & 88.22 & 79.91 & 16.75 \\ 
             &              & \checkmark   & 89.27 & 81.04 & 7.11 & 80.78 & 71.95 & 20.09 & 87.96 & 79.37 & 18.54 \\ \hline
\checkmark   & \checkmark   &              & 90.01 & 82.42 & 5.71 & 81.41 & 72.17 & 19.26 & 88.61 & 80.44 & 18.30 \\
\checkmark   &              & \checkmark   & 89.66 & 81.80 & 6.64 & 81.63 & 73.41 & 18.88 & 88.44 & 80.47 & 17.93 \\
             & \checkmark   & \checkmark   & 89.78 & 81.99 & 5.73 & 82.22 & 73.92 & 18.60 & 88.63 & 80.52 & 16.85 \\ \hline
\checkmark   & \checkmark   & \checkmark   & \textbf{90.76} & \textbf{83.48} & \textbf{4.84} & \textbf{83.03} & \textbf{74.56} & \textbf{16.99} & \textbf{89.24} & \textbf{81.67} & \textbf{15.62} \\
\hline\hline
\end{tabular}
\end{center}
\caption{Ablation of MBC, MFE, and SSA on BUS, BUSI, and GlaS. \checkmark\ denotes module inclusion. 
Baseline is a U-Net–like ETN without MBC/MFE/SSA, using standard convolutional blocks. 
Best results are in bold.}

\label{tab:ablation_components_all}
\end{table*}

\subsection{Ablation Study}
We conduct ablations to evaluate the main architectural components, loss terms, auxiliary supervision, and VAE selection. Additional sensitivity analyses are provided in Section~\ref{supp:additional_ablations}.

\subsubsection{Effect of Core Architectural Modules}
We conduct a set of experiments to evaluate the impact of our MBC, MFE, and SSA modules on segmentation performance, as shown in Table~\ref{tab:ablation_components_all}. When we use these modules independently, each generally improves segmentation quality over the baseline. Specifically, MBC alone increases BUS DSC from 88.48\% to 89.34\%, while decreasing HD95 from 6.72 to 6.53 pixels. Adding MFE raises BUSI DSC to 81.01\% and lowers GlaS HD95 to 16.75 pixels. SSA alone improves IoU across datasets. Importantly, the full combination (MBC + MFE + SSA) achieves the best overall performance, with the highest DSCs of 90.76\% (BUS), 83.03\% (BUSI), and 89.24\% (GlaS), along with the lowest HD95 values of 4.84, 16.99, and 15.62 pixels, respectively.

\subsubsection{Effect of Dilation Rate $d$ in MFE Module}
We evaluate how dilation rates affect the Multi-scale Feature Enrichment (MFE) module by comparing two- and three-branch configurations in Table~\ref{tab:dilation_ablation_combined}. The three-branch designs consistently outperform their two-branch counterparts, confirming the benefit of combining a broader range of receptive fields. Among the two-branch variants, $d={6,12}$ performs best, reaching 90.07\% DSC on BUS and 88.82\% on GlaS. Our best overall configuration, $d={1,6,12}$, combines fine, intermediate, and large receptive fields, achieving 90.76\% DSC and 4.84 HD95 on BUS, and 89.24\% DSC and 15.62 HD95 on GlaS. These results show that complementary dilation scales improve both segmentation accuracy and boundary localization.

\begin{table}[h]
\footnotesize
\begin{center}
\begin{tabular}{c|ccc|ccc}
\hline \hline
\multirow{2}{*}{Dilation Rates ($d$)} 
& \multicolumn{3}{c|}{BUS} 
& \multicolumn{3}{c}{GlaS} \\
\cline{2-7}
& DSC$\uparrow$ & IoU$\uparrow$ & HD95$\downarrow$
& DSC$\uparrow$ & IoU$\uparrow$ & HD95$\downarrow$ \\
\hline \hline
\multicolumn{7}{c}{\textbf{Two‐branch}} \\
\hline
$d= \{1,2\}$    & 89.20 & 81.26 & 6.21  & 88.21 & 78.89 & 18.83 \\
$d = \{2,4\}$    & 89.58 & 81.60 & 5.97  & 88.57 & 79.50 & 18.57 \\
$d = \{1,6\}$    & 89.82 & 81.52 & 5.38  & 88.80 & 79.83 & 18.00 \\
$d = \{6,12\}$   & 90.07 & 81.87 & 5.68  & 88.82 & 80.02 & 17.50 \\
\hline
\multicolumn{7}{c}{\textbf{Three‐branch}} \\
\hline

$d = \{1,3,9\}$  & 90.66 & 83.29 & 5.31 & 89.08 & 80.34 & 15.98 \\
$d = \{2,4,6\}$  & 90.55 & 83.15 & 5.65  & 89.10 & 80.37 & 16.03 \\
$\textbf{d = \{1,6,12\}}$ & \textbf{90.76} & \textbf{83.48} & \textbf{4.84}  
    & \textbf{89.24} & \textbf{81.67} & \textbf{15.62} \\
$d = \{1,8,12\}$ & 90.27 & 82.62 & 5.43  & 88.87 & 80.05 & 16.97 \\
$d = \{2,8,16\}$ & 90.01 & 81.83 & 5.86  & 88.88 & 80.03 & 17.60 \\
\hline \hline
\end{tabular}
\end{center}
\caption{Effect of dilation rate configurations ($d$) in the MFE module across BUS and GlaS datasets. Results are averaged over five runs.}

\label{tab:dilation_ablation_combined}
\end{table}

\subsubsection{Effect of Different Losses}
We evaluate the contribution of each loss component by adding the baseline mask-space objective $(\mathcal{L}_{\mathrm{dice}}+\mathcal{L}_{\mathrm{ftv}})$ with $\mathcal{L}_{\mathrm{rec}}$ for latent reconstruction and $\mathcal{L}_{\mathrm{la}}$ for L2 latent alignment (Table~\ref{tab:loss_ablation}). The baseline achieves 90.09\% DSC and 8.62~px HD95 on BUS, showing strong region overlap but coarse boundaries. Adding $\mathcal{L}_{\mathrm{rec}}$ improves DSC to 90.48\% and reduces HD95 to 6.40~px, indicating better latent reconstruction and boundary accuracy. Adding $\mathcal{L}_{\mathrm{la}}$ to the mask-space objective achieves 90.38\% DSC and 8.07~px HD95, providing a smaller but consistent gain. Using $\mathcal{L}_{\mathrm{rec}}+\mathcal{L}_{\mathrm{la}}$ without mask-space supervision reduces DSC to 89.60\%, highlighting the importance of segmentation guidance. The full objective achieves the best overall performance, confirming the benefit of jointly optimizing latent- and mask-space losses.

\begin{table}[t]

\footnotesize

\begin{center}
\begin{tabular}{cccc|ccc|ccc}
\hline
\multicolumn{4}{c|}{Loss} 
  & \multicolumn{3}{c|}{BUS} 
  & \multicolumn{3}{c}{GlaS} \\
\hline 
$\mathcal{L}_{\mathrm{dice}}$ & $\mathcal{L}_{\mathrm{ftv}}$ & $\mathcal{L}_{\mathrm{rec}}$ & $\mathcal{L}_{\mathrm{la}}$
  & DSC$\uparrow$ & IoU$\uparrow$ & HD95$\downarrow$ 
  & DSC$\uparrow$ & IoU$\uparrow$ & HD95$\downarrow$ \\
\hline \hline
\checkmark & \checkmark &  &  
  & 90.09 & 82.29 &  8.62 
  & 87.05 & 78.40 & 20.43 \\
\checkmark & \checkmark & \checkmark &  
  & 90.48 & 83.09 &  6.40 
  & 88.83 & 80.92 & 17.78 \\
\checkmark & \checkmark &  & \checkmark 
  & 90.38 & 82.81 &  8.07 
  & 88.05 & 79.96 & 17.81 \\ \hline
 &  & \checkmark & \checkmark 
  & 89.60 & 81.64 &  6.34 
  & 87.79 & 79.46 & 18.24 \\
\checkmark &  & \checkmark & \checkmark 
  & 90.18 & 82.58 &  5.30 
  & 88.70 & 80.87 & 17.11 \\
 & \checkmark & \checkmark & \checkmark 
  & 89.97 & 82.27 &  5.27 
  & 88.24 & 80.22 & 18.05 \\ \hline
\checkmark & \checkmark & \checkmark & \checkmark 
  & \textbf{90.76} & \textbf{83.48} & \textbf{4.84} 
  & \textbf{89.24} & \textbf{81.67} & \textbf{15.62} \\
\hline \hline
\end{tabular}
\end{center}
\caption{Ablation study of losses on BUS and GlaS. $\checkmark$ indicates the inclusion of the loss term. Bold highlights the best results.}

\label{tab:loss_ablation}

\end{table}

\subsubsection{Effect of Auxiliary Decoders at Different Stages}
To evaluate auxiliary decoding, we attach the frozen SD-VAE decoder at three ETN stages: coarse (\(s_3\)), medium (\(s_2\)), and fine (\(s_1\)), and assess both single- and multi-decoder configurations. Each auxiliary head predicts a segmentation latent, which is upsampled to the full latent resolution and decoded by the shared frozen SD-VAE decoder.

As shown in Table~\ref{tab:aux_ablation}, the fine-stage decoder (\(s_1\)) provides the best overall trade-off. Without auxiliary decoders, GET achieves \(88.51\%\) DSC / \(6.99\,\mathrm{px}\) HD95 on BUS and \(89.07\%\) DSC / \(17.14\,\mathrm{px}\) HD95 on GlaS. Adding \(s_1\) yields the strongest boundary improvement, reducing HD95 to \(4.84\,\mathrm{px}\) on BUS and \(15.62\,\mathrm{px}\) on GlaS while also improving DSC. Although \(s_1{+}s_2\) gives a slightly higher DSC on GlaS, the single \(s_1\) head provides better boundary accuracy with less redundancy. In contrast, \(s_2\) and \(s_3\) provide smaller or inconsistent gains, while multi-head configurations often introduce additional supervision noise. We therefore adopt a single \(s_1\) auxiliary decoder alongside the main output head.

\begin{table}[t]
\footnotesize
\begin{center}
\begin{tabular}{c|ccc|ccc}
\hline
\multirow{2}{*}{Aux Decoder(s)} 
& \multicolumn{3}{c|}{BUS} 
& \multicolumn{3}{c}{GlaS} \\
\cline{2-7}
& DSC$\uparrow$ & IoU$\uparrow$ & HD95$\downarrow$ 
& DSC$\uparrow$ & IoU$\uparrow$ & HD95$\downarrow$ \\
\hline \hline
None            & 88.51 & 79.93 & 6.99 & 89.07 & 81.35 & 17.14 \\
\hline
s1              & \textbf{90.76} & \textbf{83.48} & \textbf{4.84} & 89.24 & \textbf{81.67} & \textbf{15.62} \\
s2              & 90.28 & 82.60 & 5.26 & 89.18 & 81.45 & 17.41 \\
s3              & 89.79 & 82.01 & 5.61 & 88.93 & 80.96 & 17.14 \\
\hline
s1+s2           & 90.55 & 83.11 & 5.11 & \textbf{89.27} & 81.63 & 16.43 \\
s1+s3           & 89.79 & 81.97 & 5.77 & 88.56 & 80.42 & 18.28 \\
s2+s3           & 89.74 & 81.85 & 6.06 & 88.83 & 81.10 & 17.44 \\ \hline
s1+s2+s3        & 89.78 & 81.99 & 5.73 & 88.67 & 80.99 & 17.09 \\
\hline \hline
\end{tabular}
\end{center}
\caption{Effect of auxiliary decoder heads (\(s_1, s_2, s_3\)) with the main output decoder head \(out\) at different stages on BUS and GlaS datasets. Best results are in bold.}
\label{tab:aux_ablation}
\end{table}

\begin{table}[H]

\footnotesize
\begin{center}
\begin{tabular}{l|cc|cc|cc}
\hline
\multirow{2}{*}{Dataset} &
\multicolumn{2}{c|}{Dice$\uparrow$} &
\multicolumn{2}{c|}{IoU$\uparrow$} &
\multicolumn{2}{c}{HD95$\downarrow$} \\
\cline{2-7}
& SD & VQ & SD & VQ & SD & VQ \\
\hline
BUSI \cite{busi}         & \textbf{83.03} & 81.16 & \textbf{74.56} & 72.66 & \textbf{16.99} & 21.30 \\
HAM10000 \cite{ham10000}       & \textbf{94.23} & 93.09 & \textbf{89.95} & 88.19 & \textbf{9.06}  & 12.48 \\
Kvasir-Instr. \cite{kvasir_instrument} & \textbf{94.55} & 92.97 & \textbf{90.63} & 89.11 & \textbf{8.45} & 10.85 \\
\hline
\end{tabular}
\end{center}
\caption{Performance comparison on three datasets using different pre-trained VAE models. SD = SD-VAE and VQ = VQ-VAE.}
\label{tab:different_vae}
\end{table}

\subsubsection{Effect of Using Different VAE Models}
\label{sec:different_vae_models}
We compare two VAEs to identify the latent space best suited for GET. VQ-VAE~\cite{van2017neural} represents inputs using discrete vector-quantized codes, whereas SD-VAE~\cite{stable_diffusion} produces continuous, smoother embeddings. Table~\ref{tab:different_vae} shows that SD-VAE consistently achieves higher Dice and IoU and lower HD95 across all datasets. This suggests that its latent space better preserves coherent and stable representations for medical image segmentation. We therefore use SD-VAE throughout our framework.

\subsection{Computational Complexity and Efficiency}
We compare end-to-end inference cost in Table~\ref{tab:throughput}. For GET, GFLOPs and throughput include the frozen SD-VAE encoder, ETN, and decoder. GET requires 341.14 GFLOPs and achieves 28.07 samples/s, compared with 343.85 GFLOPs and 27.68 samples/s for GMS. Relative to MedSegDiff-V2, GET reduces GFLOPs by 64.9\% and achieves approximately $45\times$ higher throughput. GET also uses only 1.07M trainable parameters, compared with 1.56M for GMS, 49.84M for GSS, 129.43M for MedSegDiff-V2, and 329.17M for SDSeg.

\begin{table}[H]
\footnotesize
\begin{center}
\begin{tabular}{l|c|c|c}
\hline
Model & GFLOPs$\downarrow$ & 
\begin{tabular}[c]{@{}c@{}}Throughput\\(samples/s)$\uparrow$\end{tabular} & 
Speedup (×)$\uparrow$ \\
\hline
MedSegDiff-V2~\cite{medsegdiffv2} & 972.66 & 0.62 & $1.0\times$ \\
SDSeg~\cite{sdseg}                 & 655.34 & 15.04 & $24.3\times$ \\
GSS~\cite{gss}                     & 476.28 & 16.24 & $26.2\times$ \\
GMS~\cite{gms}                     & 343.85 & 27.68 & $44.7\times$ \\
\textbf{GET (Ours)}                & \textbf{341.14} & \textbf{28.07} & \textbf{\boldmath$45.3\times$}
 \\
\hline
\end{tabular}
\end{center}
\caption{Inference complexity of generative segmentation models at $224\times224\times3$ (batch=1) on an RTX 3090 Ti. GFLOPs are computed using THOP (MACs counted as FLOPs). Speedup ($\times$) denotes throughput relative to MedSegDiff-V2. Lower GFLOPs and higher throughput indicate better efficiency.}
\label{tab:throughput}
\end{table}
\section{Conclusion} 
We presented GET, a structured embedding-translation framework for medical image segmentation within a frozen SD-VAE space. Its multi-stage ETN combines MBC, SSA, and MFE with latent- and mask-space supervision for efficient local, global, and multi-scale modeling. Across five benchmarks, GET consistently outperforms generative, CNN, and Transformer baselines. Compared with GMS, GET improves average Dice and IoU by 0.93\% and 1.26\%, reduces HD95 by 0.81~pixels, and uses 31.41\% fewer trainable parameters. Under BUS--BUSI domain shift, it further improves Dice and IoU by 3.51\% and 3.39\% and reduces HD95 by 27.37~pixels. GET also reduces GFLOPs by 64.9\% while achieving approximately $45\times$ higher throughput than MedSegDiff-V2. These results demonstrate that structured embedding translation is accurate, robust, and efficient.

% %\clearpage  % TODO FINAL: This \clearpage needs to be removed from both review and camera-ready versions.

% \section*{Acknowledgements}
% Please insert your acknowledgments here.

\bibliographystyle{splncs04}
\bibliography{main}
\newpage
\setcounter{page}{1}
\begin{center}
    {\LARGE\bfseries Supplementary Material}
\end{center}
\vspace{1em}

In this Supplementary Material, we provide additional details on related work, evaluation metrics, and extensive ablation studies to support our architectural and design choices. We also present an efficiency–accuracy ranking plot that further demonstrates the superiority of our proposed Generative Embedding Translation (GET) framework over existing state-of-the-art (SOTA) methods.

\section{Related Work}
\label{Supp-rel}

\subsection{Convolution-Based Segmentation}
Convolutional architectures remain central to medical image segmentation. U-Net~\cite{unet} introduced the encoder--decoder design with skip connections, inspiring variants such as UNet++~\cite{unet++} with dense skip pathways and MultiResUNet~\cite{ibtehaz2020multiresunet} with multi-resolution filters. Attention U-Net~\cite{oktay2018attention} added attention gates to emphasize target regions,
while ACC-UNet~\cite{accunet} and nnU-Net~\cite{nnunet} improved hierarchical feature aggregation and automated configuration. Lightweight designs such as AULUNet~\cite{aulunet}, EGE-UNet~\cite{egeunet}, and LB-UNet~\cite{lbunet} reduce computation while preserving accuracy. Despite their popularity, CNNs rely on local receptive fields, which limit their ability to capture global structure that is often crucial for complex or ambiguous anatomy.

\subsection{Transformer-Based Segmentation}
Transformers address this limitation through global self-attention~\cite{vit}. Hybrid models such as TransUNet~\cite{transunet} and MedT~\cite{medt} combine CNN encoders with transformer bottlenecks to capture both local and global dependencies. Fully transformer-based designs, including SwinUNet~\cite{swinunet}, MISSFormer~\cite{missformer}, and UNETR~\cite{unetr}, stack hierarchical transformer blocks with windowed attention to maintain large receptive fields at moderate cost. These methods improve long-range reasoning but still require high computational overhead due to quadratic (or near-quadratic) attention and often require large annotated datasets, which makes deployment difficult in data- and resource-limited clinical settings~\cite{mambaliteunet,  trmmedicalsurvey,gms}. These limitations motivate exploring alternatives \cite{mambainvision, gss, gms} that leverage rich semantic priors without adding the computational burden of heavy transformer backbones.

\subsection{Generative Models for Segmentation}
Generative models provide latent semantic priors that can enhance robustness and data efficiency. VAEs~\cite{iso_kl,nave} and VQ-VAEs~\cite{van2017neural,taming} learn structured latent spaces that support downstream segmentation by providing smoother and more semantically organized representations than raw pixels. Diffusion-based methods such as MedSegDiff-V2~\cite{medsegdiffv2} and SDSeg~\cite{sdseg} further leverage strong generative priors and achieve high-quality masks through iterative denoising, but they are computationally expensive and slow at inference.
Latent-space approaches such as GSS~\cite{gss} and GMS~\cite{gms} avoid iterative denoising by directly operating on pretrained latent representations. GMS maps image latents to mask latents using a pretrained VAE, demonstrating the feasibility of direct image-to-mask latent translation. GET operates within this broader latent-translation paradigm but focuses on a structured, multi-stage transformation. Its ETN combines efficient local modeling, global context aggregation, and multi-scale refinement, together with latent- and mask-space supervision, to preserve spatial structure while keeping the trainable component compact.

\section{Experiments and Results}
\label{supp:exp_res}
This section extends our “Experiments and Results Section” with evaluation details, results comparison, and additional experiments.

\subsection{Evaluation Metrics}
\label{supp:eval_metrics}
We evaluate segmentation performance using the Dice Similarity Coefficient (DSC), Intersection over Union (IoU), and the 95th-percentile Hausdorff Distance (HD95) to measure the overall segmentation performance of the models on all datasets. These metrics collectively evaluate overlap (DSC), spatial accuracy (IoU), and boundary precision (HD95). Together, they offer a comprehensive evaluation of segmentation quality. The metrics are defined as:
\begin{equation}
\mathrm{DSC}(M, \widehat{S}) = \frac{2\,|M \cap \widehat{S}|}{|M| + |\widehat{S}|} \times 100,
\label{eq:dice}
\end{equation}
\begin{equation}
\mathrm{IoU}(M, \widehat{S}) = \frac{|M \cap \widehat{S}|}{|M \cup \widehat{S}|} \times 100,
\label{eq:iou}
\end{equation}
\begin{equation}
\begin{aligned}
\mathrm{HD95}(M, \widehat{S}) = \max\Big(
&\mathrm{P}_{95}\big( \min_{\hat{s} \in \widehat{S}} d(m, \hat{s}) \big)_{m \in M}, \\
&\mathrm{P}_{95}\big( \min_{m \in M} d(\hat{s}, m) \big)_{\hat{s} \in \widehat{S}}
\Big),
\label{eq:hd95}
\end{aligned}
\end{equation}
where $M$ and $\widehat{S}$ denote the sets of boundary points of the ground-truth and generated segmentation masks, $d(\cdot,\cdot)$ measures the Euclidean distance between boundary points, and $\mathrm{P}_{95}(\cdot)$ denotes the 95th percentile over a set of distances. HD95 measures boundary alignment accuracy as the 95th percentile of distances between points on the predicted and ground-truth boundaries, which makes it robust to outliers.

\subsection{Training Loss}
\label{supp:train_loss_formulas}

For training, we optimize the hybrid objective:

\begin{equation}
L_{\text{total}} = L_{\text{rec}} + L_{\text{la}} + L_{\text{seg}}.
% \tag{17}
\end{equation}

Here, the latent reconstruction term $\mathcal{L}_{\mathrm{rec}}$ and the L2 latent alignment term $\mathcal{L}_{\mathrm{la}}$ are already defined in Eqs.~10 and 11, respectively, in the main paper. The mask-space supervision $L_{\text{seg}}$ is given in Eq.~\ref{eq:main_lseg}.
For completeness, we restate the overlap-based losses (Dice loss $L_{\text{dice}}$ and Focal Tversky loss $L_{\text{ftv}}$) that form $L_{\text{seg}}$. $L_{\text{dice}}$ and $L_{\text{ftv}}$ are used to address class imbalance and sharpen boundary predictions.
We define them as:
\begin{equation}
  \mathcal{L}_{\mathrm{dice}}(M,\widehat{S})
  = 1 - \frac{2\,|M\cap\widehat{S}|}
               {|M| + |\widehat{S}|}\,,
  \label{eq:dice_loss}
\end{equation}

\begin{equation}
  \mathcal{L}_{\mathrm{ftv}}(M,\widehat{S})
  = \Biggl[1 - 
      \frac{|M\cap\widehat{S}| + \epsilon}
           {|M\cap\widehat{S}| 
            + \alpha\,|M\setminus\widehat{S}| 
            + \beta\,|\widehat{S}\setminus M| 
            + \epsilon}
  \Biggr]^{\gamma},
  \label{eq:ftv_loss}
\end{equation}
where $\epsilon = 10^{-6}$ prevents division by zero, and standard hyperparameters $\alpha = 0.7$, $\beta = 0.3$, and $\gamma = 0.75$ balance sensitivity to false positives and negatives while focusing on hard examples \cite{ftvloss}, and $M$ and $\widehat{S}$ denote the ground-truth and generated segmentation masks.

\subsection{Model Efficiency-Accuracy Trade-off}
\label{supp:model_eff_trade-off}

\begin{figure*}[t]
  \begin{center}
  \includegraphics[width=\textwidth]{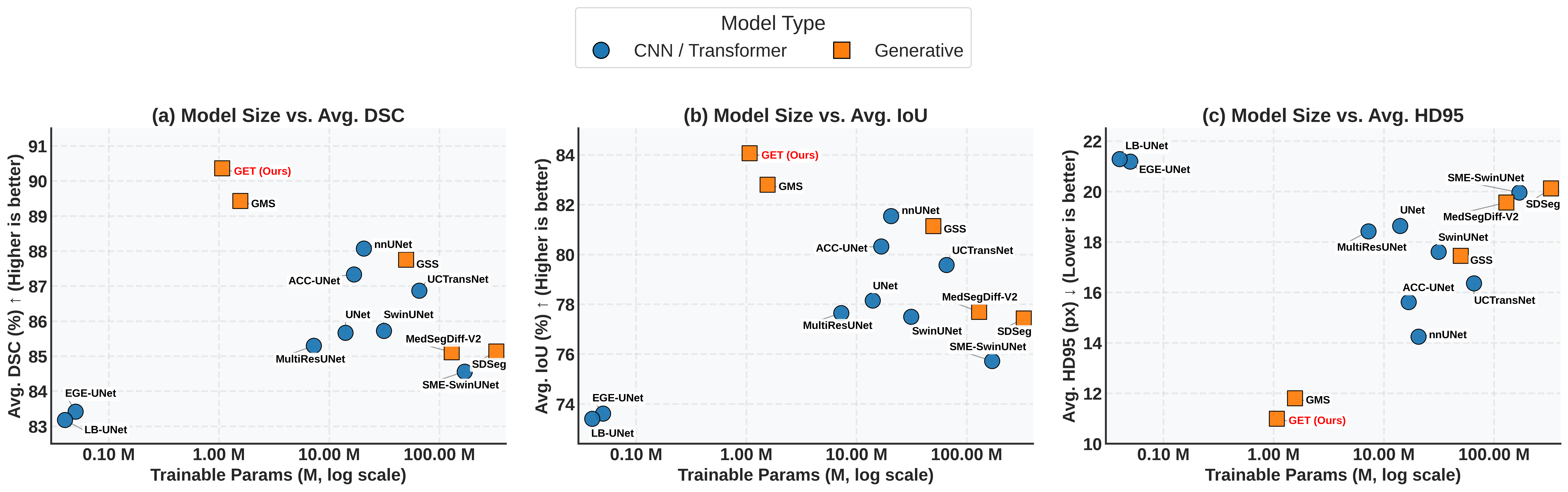}
  \end{center}
\caption{Comparison of segmentation performance versus trainable parameter count across SOTA methods, averaged over five medical imaging datasets: BUS, BUSI, GlaS, HAM10000, and Kvasir--Instrument. (a) Trainable parameters vs. average DSC score. (b) Trainable parameters vs. average IoU score. (c) Trainable parameters vs. average HD95 score. Higher DSC and IoU ($\uparrow$) indicate better overlap accuracy, while lower HD95 ($\downarrow$) reflects better boundary precision. Blue circles denote CNN/Transformer-based models, and orange squares denote generative models. GET (Ours) achieves a favorable efficiency--accuracy trade-off, with top-tier DSC, IoU, and HD95 performance using substantially fewer trainable parameters than existing SOTA methods.}
  \label{fig:ranking}
\end{figure*}

To evaluate the trade-off between segmentation accuracy and model complexity, we present a comparative ranking plot in Figure~\ref{fig:ranking}, which visualizes the relationship between trainable parameters and segmentation performance. Specifically, we plot the average Dice Similarity Coefficient (DSC), Intersection over Union (IoU), and 95th percentile Hausdorff Distance (HD95) against the number of trainable parameters on a log scale, aggregated across five benchmark datasets: BUS~\cite{bus}, BUSI~\cite{busi}, GlaS~\cite{glas}, HAM10000~\cite{ham10000}, and Kvasir--Instrument~\cite{kvasir_instrument}.

Models positioned closer to the top-left regions in the DSC and IoU plots and the bottom-left region in the HD95 plot are considered better performers, as they provide high segmentation accuracy with low model complexity. As shown in Figure~\ref{fig:ranking}, our GET model consistently outperforms recent state-of-the-art approaches, including CNN-, Transformer-, and diffusion-based generative methods. GET achieves superior DSC and IoU scores and lower HD95 while significantly reducing parameter overhead. Therefore, GET offers a strong balance between representational power and efficiency, making it well-suited for medical image segmentation scenarios where both precision and lightweight design are critical.

\section{Additional Ablation Study}
\label{supp:additional_ablations}

To further demonstrate the robustness and sensitivity of our GET framework, we conduct extensive ablation studies on critical parameters and architectural choices using two datasets: BUS~\cite{bus} (ultrasound) and GlaS~\cite{glas} (histology). These experiments validate the individual contribution of each component to the model’s overall performance.

\subsection{Effect of Subsampling Factor $r$ in SSA Module}
\label{supp:r_effect}

\begin{table}[t]
\small
\setlength{\tabcolsep}{1.5mm}
\begin{center}
\begin{tabular}{c|ccc|ccc}
\hline
\multirow{2}{*}{$SSA(r)$} 
& \multicolumn{3}{c|}{BUS} 
& \multicolumn{3}{c}{GlaS} \\
\cline{2-7}
& DSC$\uparrow$ & IoU$\uparrow$ & HD95$\downarrow$ 
& DSC$\uparrow$ & IoU$\uparrow$ & HD95$\downarrow$ \\
\hline
1 & 90.40 & 82.89 & 5.19 & 89.03 & 81.27 & 16.42 \\
\textbf{2} & \textbf{90.76} & \textbf{83.48} & \textbf{4.84} & \textbf{89.24} & \textbf{81.67} & \textbf{15.62} \\
4 & 88.72 & 80.53 & 6.45 & 87.63 & 79.44 & 18.18 \\
\hline
\end{tabular}
\end{center}
\caption{Effect of subsampling factor $r$ in the SSA module on BUS and GlaS datasets. A moderate setting ($r=2$) yields the best trade-off between segmentation accuracy (DSC and IoU) and boundary precision (HD95). Results are averaged over five runs; best values are in bold. $\uparrow$ indicates higher is better, $\downarrow$ lower is better.}

\label{tab:ssa_r_ablation}
\end{table}

We evaluate the impact of subsampling factor ($r$) in our Subsampled Self-Attention (SSA) module. 
Table~\ref{tab:ssa_r_ablation} presents our experimental results.
% , which reduces computational overhead by downsampling spatial dimensions in self-attention.
From the table, we can see that the moderate subsampling factor ($r=2$) achieves the best balance, gaining peak performance on both datasets (90.76\% DSC on BUS and 89.24\% DSC on GlaS) and lowest HD95 values. In contrast, excessive subsampling ($r=4$) considerably reduces accuracy due to lost spatial detail, while no subsampling ($r=1$) introduces additional computational complexity without improved results. Thus, $r=2$ provides an optimal efficiency-accuracy trade-off.

\subsection{Effect of Deep Supervision (DS) Weights %
\texorpdfstring{$w_k$}{wk} on Auxiliary Heads for %
\texorpdfstring{$\mathcal{L}_{\mathrm{rec}}$}{Lrec} Loss}

\label{supp:ds_effect}

\begin{table}[t]
\footnotesize
\setlength{\tabcolsep}{0.68mm}
\begin{center}
\begin{tabular}{p{2.85cm}|ccc|ccc}
\hline
\multirow{2}{*}{DS Weights ($w_k$)} 
& \multicolumn{3}{c|}{BUS} 
& \multicolumn{3}{c}{GlaS} \\
\cline{2-7}
& DSC$\uparrow$ & IoU$\uparrow$ & HD95$\downarrow$
& DSC$\uparrow$ & IoU$\uparrow$ & HD95$\downarrow$ \\
\hline \hline
\makecell[l]{Uniform weights:\\(1.0, 1.0, 1.0, 1.0)}
& 86.42 & 78.72 & 11.62 & 86.37 & 77.37 & 21.82 \\ \hline
\makecell[l]{Linear weights:\\(1.0, 0.75, 0.5, 0.25)}
& 89.22 & 81.43 & 10.65 & 88.40 & 80.48 & 16.79 \\ \hline
\makecell[l]{Monotonic weights:\\\textbf{(1.0, 0.4, 0.3, 0.2)}}
& \textbf{90.76} & \textbf{83.48} & \textbf{4.84} 
& \textbf{89.24} & \textbf{81.67} & \textbf{15.62} \\
\hline \hline
\end{tabular}
\end{center}
\caption{Effect of deep supervision weighting strategies ($w_k$) on auxiliary heads for $\mathcal{L}_{\mathrm{rec}}$ loss across BUS and GlaS datasets. Results are averaged over five runs. $\uparrow$ indicates higher is better, $\downarrow$ lower is better. Best values are highlighted in bold.}
\label{tab:ds_weights_ablation}
\end{table}

We next examine the deep supervision weighting strategy ($w_k$) applied to multiple auxiliary heads for $\mathcal{L}_{\mathrm{rec}}$ loss (Eq.~\ref{eq:main_lrec} in the manuscript). We compare uniform, linear, and monotonically decaying weighting schemes to identify the best approach. 

Our results are summarized in Table~\ref{tab:ds_weights_ablation}.
We found that monotonically decaying weights ($w_k=\{1.0, 0.4, 0.3, 0.2\}$) consistently outperform both uniform and linear approaches, achieving the highest DSC (90.76\% on BUS, 89.24\% on GlaS) and lowest HD95.
This indicates that the monotonic scheme effectively emphasizes fine-scale features critical for boundary accuracy while still benefiting from coarse-to-fine guidance, validating our chosen weighting strategy.

\subsection{Effect of Auxiliary Head Weight %
\texorpdfstring{$\lambda$}{lambda} in the %
\texorpdfstring{$\mathcal{L}_{\mathrm{seg}}$}{Lseg} Loss}

\label{supp:aux_head_effect}

\begin{table}[t]
\small
\begin{center}
\begin{tabular}{c|ccc|ccc}
\hline 
\multirow{2}{*}{$\lambda$} 
& \multicolumn{3}{c|}{BUS} 
& \multicolumn{3}{c}{GlaS} \\
\cline{2-7}
& DSC$\uparrow$ & IoU$\uparrow$ & HD95$\downarrow$ 
& DSC$\uparrow$ & IoU$\uparrow$ & HD95$\downarrow$ \\
\hline
0.1 & 89.83 & 81.55 & 7.15  & 87.86 & 79.51 & 17.26 \\
0.3 & 90.12 & 82.41 & 5.94  & 88.15 & 80.00 & 17.49 \\
0.4 & 90.40 & 82.89 & 5.19  & 88.63 & 80.57 & 18.50 \\
\textbf{0.5} & \textbf{90.76} & \textbf{83.48} & \textbf{4.84}  
             & \textbf{89.24} & \textbf{81.67} & \textbf{15.62} \\
0.6 & 90.28 & 82.73 & 5.76  & 89.06 & 81.12 & 16.04 \\
0.7 & 90.02 & 82.33 & 5.90  & 88.43 & 80.56 & 18.27 \\
\hline
\end{tabular}
\end{center}
\caption{Effect of auxiliary head weight ($\lambda$) in $\mathcal{L}_{\mathrm{seg}}$ loss across BUS and GlaS datasets. The weight $\lambda$ controls the contribution of intermediate supervision from the auxiliary decoder head. Optimal performance is achieved at $\lambda=0.5$, balancing guidance strength and segmentation accuracy. $\uparrow$ indicates higher is better, $\downarrow$ lower is better. Results are averaged over five runs; best values are highlighted in bold.}
\label{tab:lambda_ablation}
\end{table}

We analyze the influence of the auxiliary-head loss weight ($\lambda$) in %
$\mathcal{L}_{\mathrm{seg}}$ (in manuscript: Eq.~\ref{eq:main_lseg}), which determines how strongly the intermediate supervision from decoder stage $s1$ guides segmentation. 

Our results in Table~\ref{tab:lambda_ablation} show that $\lambda = 0.5$ achieves the best balance: it yields the highest DSC (90.76\% on BUS, 89.24\% on GlaS) and the lowest boundary errors (HD95 of 4.84 and 15.62). Smaller weights ($\lambda<0.5$) provide insufficient guidance, while larger ones ($\lambda>0.5$) introduce conflicting gradients that degrade accuracy. Thus, $\lambda = 0.5$ offers the optimal trade-off between intermediate supervision strength and final segmentation quality.

\subsection{Effect of L2 Latent Alignment Weight $\lambda_{\mathrm{la}}$ in the \texorpdfstring{$\mathcal{L}_{\mathrm{la}}$}{L_la} Loss}

\label{supp:la_effect}

\begin{table}[t]
\small
\setlength{\tabcolsep}{1.2mm}
\begin{center}
\begin{tabular}{c|ccc|ccc}
\hline
\multirow{2}{*}{$\lambda_{la}$} 
& \multicolumn{3}{c|}{BUS} 
& \multicolumn{3}{c}{GlaS} \\
\cline{2-7}
& DSC$\uparrow$ & IoU$\uparrow$ & HD95$\downarrow$ 
& DSC$\uparrow$ & IoU$\uparrow$ & HD95$\downarrow$ \\
\hline
0.001 & 88.52 & 80.34 & 8.94  & 88.37 & 80.19 & 17.89 \\
0.005 & 89.25 & 81.41 & 7.78  & 88.94 & 80.97 & 16.38 \\
\textbf{0.01} & \textbf{90.76} & \textbf{83.48} & \textbf{4.84}
              & \textbf{89.24} & \textbf{81.67} & \textbf{15.62} \\
0.02  & 89.02 & 81.04 & 6.94  & 88.72 & 80.89 & 17.06 \\
0.05  & 88.13 & 79.93 & 9.05  & 88.21 & 79.97 & 17.14 \\
0.1   & 87.44 & 78.72 & 10.40 & 87.75 & 79.44 & 18.05 \\
\hline
\end{tabular}
\end{center}
\caption{Effect of the L2 latent alignment weight $\lambda_{\mathrm{la}}$ in $\mathcal{L}_{\mathrm{la}}$ across BUS and GlaS. The best performance is obtained at $\lambda_{\mathrm{la}}=0.01$. Results are averaged over five runs; best values are highlighted in bold. $\uparrow$ indicates higher is better and $\downarrow$ lower is better.}
\label{tab:la_ablation_bus_glas}
\end{table}

Lastly, we study the effect of the L2 latent alignment weight $\lambda_{\mathrm{la}}$ in $\mathcal{L}_{\mathrm{la}}$ (Eq.~11). Table~\ref{tab:la_ablation_bus_glas} shows that $\lambda_{\mathrm{la}}=0.01$ provides the best overall performance, achieving 90.76\% DSC on BUS and 89.24\% DSC on GlaS, with HD95 values of 4.84 and 15.62 pixels, respectively. Lower or higher weights result in weaker segmentation performance, indicating that a moderate L2 alignment constraint provides the best balance between latent-space supervision and segmentation accuracy.
\end{document}